\documentclass[]{aastex631}

\begin{document}

\title{Neutrinos and gamma-rays from Galaxy Clusters constrained by the upper limits of IceCube}

\author[0000-0002-0458-0490]{Saqib Hussain}
\affiliation{Gran Sasso Science Institute, Via Michele Iacobucci, 2, 67100 L'Aquila AQ, Italy}
\affiliation{Laboratori Nazionali del Gran Sasso - LNGS - INFN, Italy}


\author[0000-0001-8058-4752]{Elisabete M. {de Gouveia Dal Pino}}
\affiliation{Institute of Astronomy, Geophysics and Atmospheric Sciences (IAG), University of S\~ao Paulo (USP), S\~ao Paulo, Brazil}

\author[0000-0002-6751-9996]{Giulia Pagliaroli}
\affiliation{Gran Sasso Science Institute, Via Michele Iacobucci, 2, 67100 L'Aquila AQ, Italy}
\affiliation{Laboratori Nazionali del Gran Sasso - LNGS - INFN, Italy}








\begin{abstract}
Clusters of galaxies possess the capability to accelerate cosmic rays (CRs)  to very high energy up to $\sim10^{18}$~eV due to their large size and magnetic field strength which favor CR confinement for cosmological times.
During their confinement, they can produce neutrinos and $\gamma-$rays out of interactions with the  background gas and photon fields.
In recent work, \cite{hussain2021high, hussain2023diffuse} have 
conducted three-dimensional cosmological magnetohydrodynamical (MHD) simulations of the turbulent intracluster medium (ICM) combined with multi-dimensional Monte Carlo simulations of CR propagation for redshifts ranging from $z \sim 5$ to $z = 0$ to study the multi-messenger emission from these sources.  They found that when CRs with a spectral index in the range $1.5 - 2.5$ and cutoff energy $E_\mathrm{max} = 10^{16} - 10^{17}$~eV are injected into the system, 
they make significant contributions to the diffuse background emission of both neutrinos and gamma-rays. 
In this work, we have revisited this model  
and undertaken further constraints on the parametric space. This was achieved by incorporating the recently established upper limits on neutrino emission from galaxy clusters, as obtained by the IceCube experiment.
We find that for CRs injected with spectral indices in the range  $2.0 - 2.5$, cutoff energy $E_\mathrm{max} = 10^{16} - 10^{17}$~eV, 
and power corresponding to  $(0.1-1)\%$ of the cluster luminosity,
our neutrino flux aligns with the upper limits estimated by IceCube.
Additionally, the resulting contribution from clusters to the diffuse  $\gamma$-ray background (DGRB) remains significant with values of the order of $ \sim 10^{-5}\, \mathrm{MeV} \, \mathrm{cm}^{-2} \,\mathrm{s}^{-1} \, \mathrm{sr}^{-1}$  at energies above $500$ GeV.


\end{abstract}

\keywords{Multi-messenger Astronomy; Galaxy Clusters}


\section{Introduction} \label{sec:intro}

The diffuse neutrino \citep{aartsen2015searches} and gamma ray \citep{ackermann2015spectrum} backgrounds provide a unique prospective of the high-energy Universe but their origin is still debated. It is
closely related with ultra-high-energy cosmic rays (UHECRs) \citep{abu2013cosmic, aab2017combined}.
The most plausible scenario is that the neutrino and gamma-rays are produced  by interactions of UHECRs with   the background gas and photon fields in astrophysical environments.
Several sources embedded in galaxy clusters arise as candidates for the production of very high energy CRs including active galactic nuclei and starburst galaxies. The  turbulent intracluster medium (ICM)  is also believed to be particularly suitable to accelerate and confine CRs up to energies $\sim 10^{18}$~eV \citep{inoue2007ultrahigh, wiener2019constraints, batista2019cosmogenic}.
%

Recently, the IceCube collaboration \citep{abbasi2022searching} performed a stacking analysis for $1094$ galaxy clusters of masses $\gtrsim 10^{14}\, M_{\odot}$ and 
up to redshift $z\lesssim  1.0$ 
taken from the  $2015$ PLANCK survey  \citep{ade2016planck}.
%
To complete the catalogue, 
they calculated the distribution of galaxy clusters by drawing $\sim 10^5$ samples with mass range  $10^{14} \lesssim M/M_{\odot}\lesssim 10^{15}$ extending to redshift $z\lesssim 2.0$, using  the Tinker$-2010$ halo mass function \citep{tinker2010large}. 
Based on these samples, they estimated  upper limits according to which the contribution from the clusters to the observed diffuse neutrino background is at most $\sim 4.6\%$ at 100 TeV 
for the most realistic scenario \cite[see for details][]{raghunathan2022constraining}.

Several recent studies \cite[see e.g., ][]{murase2013testing,fang2016high, fang2018linking, hussain2021high, hussain2023diffuse}  have predicted that clusters of galaxies can contribute to a sizeable percentage to the diffuse neutrino and $\gamma-$ray background.
In particular,  \cite{hussain2021high, hussain2023diffuse} used the most detailed numerical approach to date combining three-dimensional (3D) cosmological magneto-hydro-dynamic (MHD) simulations with Monte-Carlo simulations of CR propagation and cascading and predicted that clusters can potentially contribute to a fairly large fraction
to the diffuse neutrino and gamma-ray background observed by IceCube \citep{aartsen2015searches} and Fermi-LAT \citep{ackermann2015spectrum}, respectively. 
However, the aforementioned upper limit reported by the IceCube \citep{abbasi2022searching} excludes part of the parametric space they considered in their analysis of the neutrino flux. 
%

In this work, we constrain the parametric space considered in  \cite{hussain2021high, hussain2023diffuse} employing the latest upper limits reported by the IceCube for the neutrino flux of galaxy clusters and derive new limits to the diffuse $\gamma-$ray flux. We find that these new constraints eliminate the harder CR spectral indices  $\alpha \lesssim 2.0$, but still predict a substantial contribution from the clusters to the very high energy range of the  $\gamma-$ray flux.

In section~\ref{sec:method} we describe the methodology for the calculation of the fluxes of neutrinos and $\gamma-$rays in galaxy clusters, in section~\ref{sec:results} we present the results, and   section~\ref{sec:conclusion} is dedicated to the discussion of our results and the conclusions.




\section{Methodology}\label{sec:method}

We have employed here the same set of simulations as in  \cite{hussain2021high, hussain2023diffuse}.
The  ICM was modeled with  3D-MHD cosmological simulations \citep{dolag2005constrained} up to redshift $z\lesssim 5.0$ taking into account the non-uniform distribution evolution of the magnetic field, gas density, and temperature. 
CRs were proppagated in the ICM and intergalactic medium (IGM) 
to produce neutrinos and $\gamma-$rays, employing the Monte Carlo code CRpropa \citep{batista2016crpropa}.
To calculate the neutrino flux  CRs were injected in the energy range  $10^{14} \leq E/\mathrm{eV}\leq 10^{19}$ because we are interested in neutrino energies above TeV. 
On the other hand,  for  $\gamma-$rays,  since  we are interested  in energies above $10$~GeV,  CRs were injected with energies $10^{11}\leq E/\mathrm{eV} \leq 10^{19}$.
However, in order to normalize the total energy to the luminosity of the clusters, it was considered the whole energy range starting from $1$~GeV.
Also, in order to account for different sources of acceleration,
 CRs were injected at different locations within the clusters: in the center, at a radius $\sim 300$~kpc, and in the outskirt $\sim 1$~Mpc.
Obviously, the dominant contribution to $\gamma-$ray and neutrino fluxes come from CRs sources located in the center
\citep[see][for details]{hussain2021high, hussain2023diffuse}.
In these previous studies,
it has been  considered that $1\%$ of the cluster luminosity ($L_C$) goes to the CRs.  

The simulations have two steps. In the first step,  CRs are propagated inside the clusters and $\gamma-$rays and neutrinos are collected at the edge of them.
All relevant CR interactions were considered during their propagation inside clusters, namely,
 proton-proton (pp) interactions, photopion production, Bethe-Heitler pair production, pair production, and inverse Compton scattering (ICS). 
Energy losses due to the expansion of the universe and the synchrotron emission were also considered, but the photons produced in these processes are below the energy range of interest of this work.
In the second step, the $\gamma-$rays collected at the boundary of the clusters were propagated through the intergalactic medium (IGM) across the redshift interval.
%
During this propagation,  the electromagnetic cascade processes including (single, double, and triplet) pair production and ICS were also accounted for.
The effect of the IGM magnetic field was neglected in this propagation step since it does not significantly affect the $\gamma-$ray flux above $10$~GeV \citep{hussain2023diffuse}.
%
%

As in \cite{hussain2021high} and \cite{hussain2023diffuse},  the  integrated neutrino and $\gamma-$ray fluxes ($\Phi$) from all clusters in the  mass range $10^{12} \lesssim M/M_{\odot}\lesssim 2\times 10^{15}$ and $z \leq 5.0$, are obtained from:
\begin{equation}
E_\mathrm{obs}^{2} \Phi(E_\mathrm{obs}) =  \int\limits_{z_\mathrm{min}}^{z_\mathrm{max}} dz \int\limits_{M_\mathrm{min}}^{M_\mathrm{max}} dM \frac{dN}{dM} E^{2}  \frac{d\dot{N}( E/(1+z), M , z)}{dE}  \times
g(E_\mathrm{obs}, E, z)
\left(\frac{\psi_{\mathrm{ev}}(z) f(M)}{4\pi d_L^2 (z)}\right) 
\end{equation} 
where $dN/dM$ is the number of clusters per mass interval calculated from the MHD simulation, $g(E_\mathrm{obs}, E, z)$
accounts for the interactions of gamma rays in the ICM and the IGM,
$\psi_{\mathrm{ev}}(z)$ is a function that describes the cosmological evolution of the emissivity of the CR sources \citep[see e.g.,][]{batista2019cosmogenic}, the quantity $E^2 \; d\dot{N}/dE$ denotes the neutrino or $\gamma-$ray power spectrum obtained from the simulations,  $d_L$ is the luminosity distance, and $f(M)$ is a factor that accounts for stellar and AGN feedback \citep{planelles2014role}.
For detailed calculations, we refer to \cite{hussain2023diffuse} and \cite{hussain2021high}.

\section{Results}\label{sec:results}



\begin{figure}[ht!]
\centering
\includegraphics[width=0.7\textwidth]{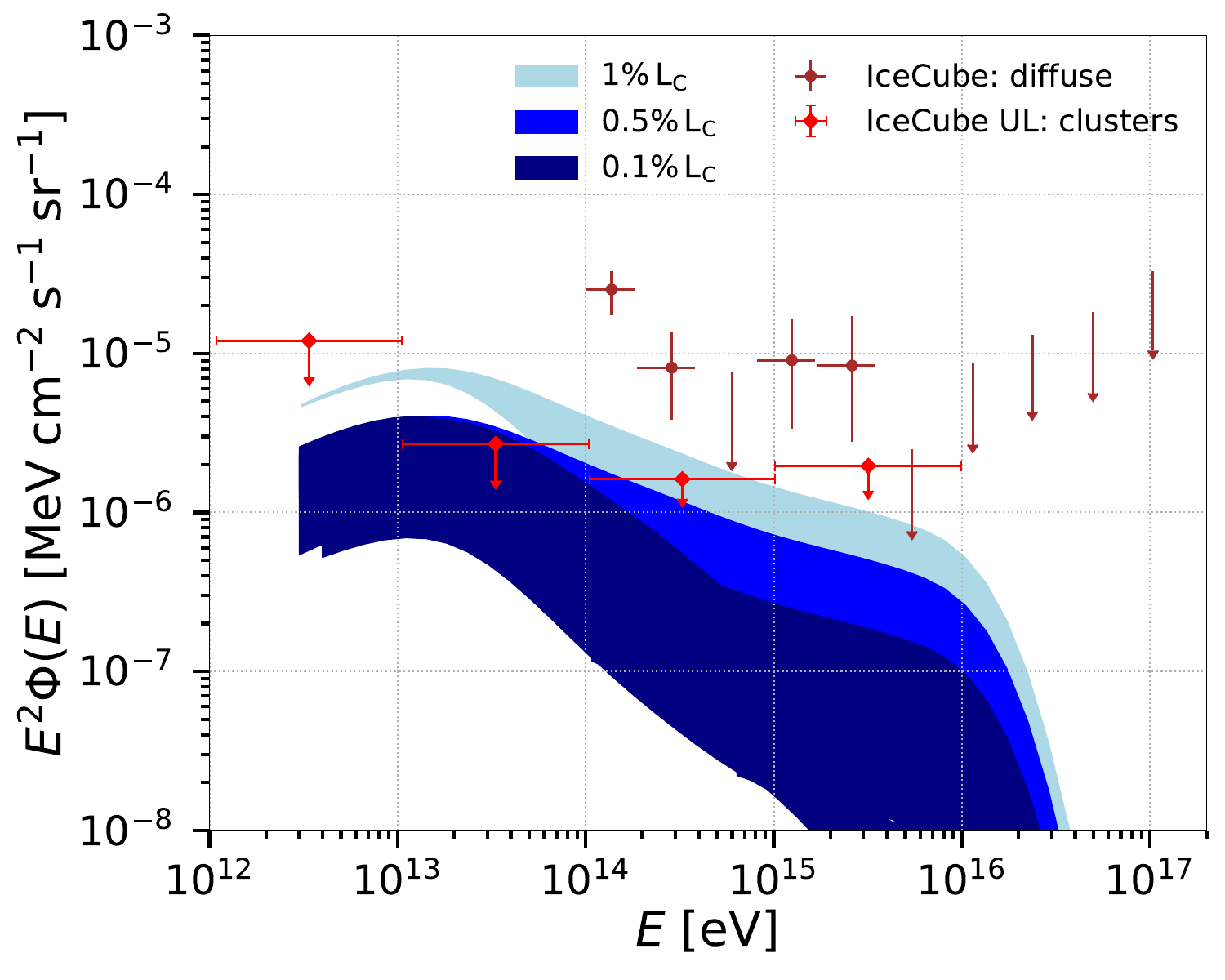}
\caption{Neutrino flux from the entire population of clusters up to redshift $z\leq 5.0$. From top to bottom blue bands correspond to $1\%, \, 0.5\%$, and $0.1\% $ of the
cluster luminosity ($L_C$), respectively. Light-blue and blue bands represent the flux for $2.0\leq \alpha \leq 2.5$ and $E_\mathrm{max}=10^{17}$~eV. The dark-blue band shows the flux for the same spectral indices, but  $E_\mathrm{max}$ ranges from $10^{16} - 10^{17}$~eV.
The IceCube upper limits for clusters \citep{abbasi2022searching} as well as the diffuse neutrino background  \citep{aartsen2015evidence, aartsen2015searches} are also depicted. 
}
\label{fig:UL_Cluster}
\end{figure}

In Fig. \ref{fig:UL_Cluster}, we show the flux of  neutrinos from clusters of galaxies,
considering that their contribution 
is constrained by the upper limits recently reported  by the  IceCube \citep{abbasi2022searching}.
As stressed, in the previous works \cite{hussain2021high,hussain2023diffuse} had assumed that  $1\%$ of each cluster luminosity ($L_C$) goes to CRs,  
and considered a range for the CR spectral index $1.5\leq \alpha \leq2.5$ and maximum energy  $10^{16}\leq E_\mathrm{max}/\mathrm{eV} \leq 5\times 10^{17}$. 
Here, we find that the CR parameters which are more suitable to fulfill the new IceCube limits
are the following: $2.0 \leq \alpha \leq 2.5$,
$E_{max} = 10^{16} - 10^{17}$~eV, and luminosities in the range $(0.1-1.0)\% \, L_C$. 
The figure shows three bands all constrained by this interval of $\alpha$ values, but with three different luminosities  $1\% L_C, \, 0.5\% L_C$, and $0.1\% L_C$, from light to dark-blue, respectively. Also, while the light-blue and blue bands have  $E_\mathrm{max}=10^{17}$~eV, the dark-blue band shows the flux for   $E_\mathrm{max}$ in the range  $10^{16} - 10^{17}$~eV. In each of these bands, the larger the value of the index $\alpha$, the smaller the flux is. 
We see that the  band with  $0.1\%$ $L_C$ is the most constrained one by the IceCube limits. In particular, for $\alpha \geq 2.3$, 
this band falls entirely below the IceCube limits.
Also,  decreasing the value of $E_{max}$ results in larger flux  at  smaller neutrino energies. This explains why the 
light-blue and blue bands produce fluxes around or below the IceCube limits only for $E_{max} \simeq 10^{17}$~eV. For $E_{max} \simeq 10^{16}$~eV, the peak of the flux in the figure increases by almost an order of magnitude at neutrino energies $\sim 10^{13}$ eV.

\begin{figure}[ht!]
\centering
\includegraphics[width=0.7\textwidth]{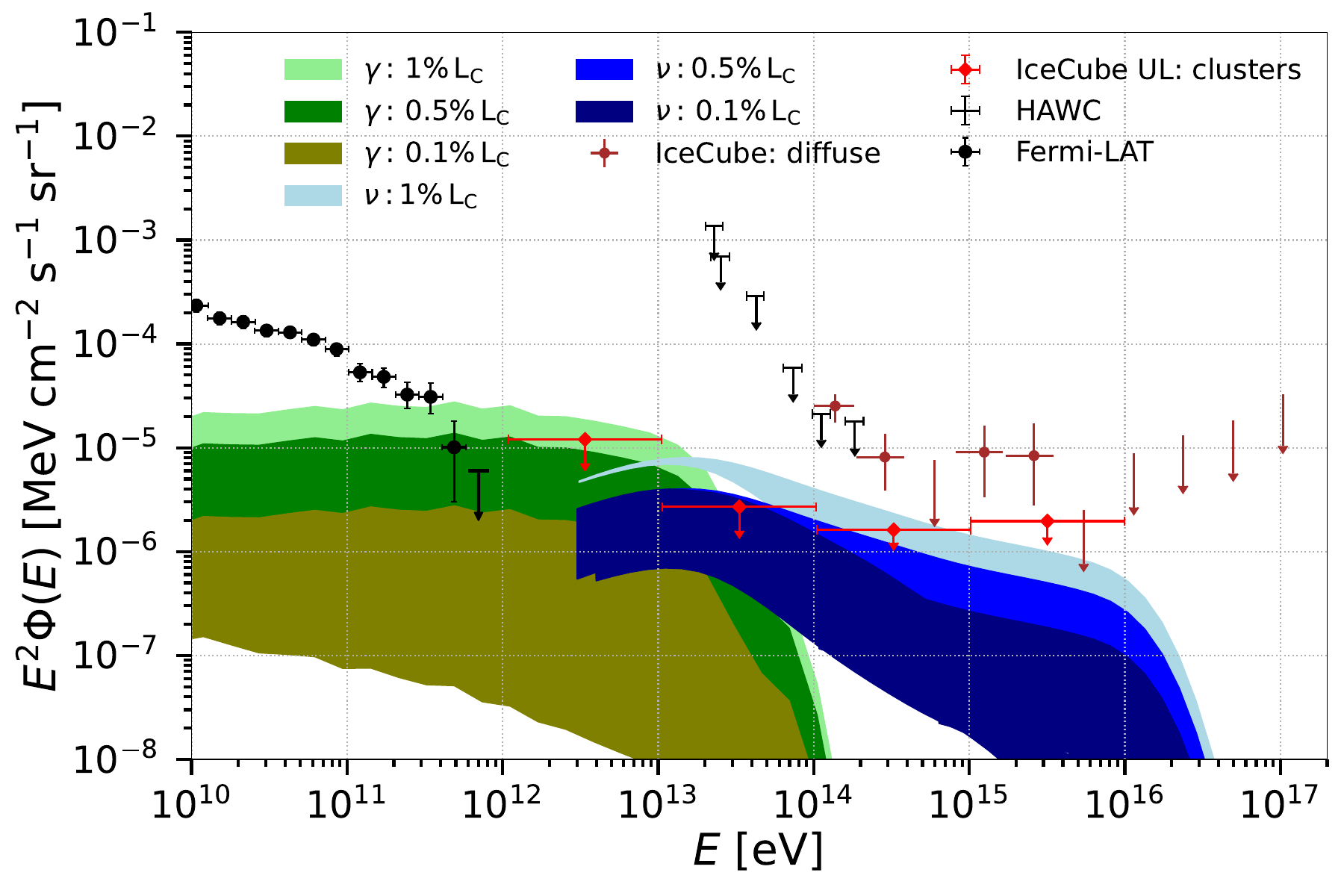}
\caption{Multi-messenger emission from clusters of galaxies.  
Blue bands are the same as presented in Fig.~\ref{fig:UL_Cluster} for neutrinos. The green bands give the diffuse flux of $\gamma-$ray obtained for the same CR parametric space as  in Fig.~\ref{fig:UL_Cluster} that suits the IceCube upper limits. The $\gamma-$ray fluxes in the light-green, dark-green and olive-green bands, have the same parameters as  the neutrino fluxes in the light-blue, blue and dark-blue, respectively.
The diffuse neutrino background upper limits reported by the IceCube \citep{aartsen2015searches, aartsen2015evidence}, 
the DGRB observed by Fermi-LAT~\citep{ackermann2015spectrum}, and the upper limits for the DGRB from HAWC \citep{HAWC2022limits} are also depicted. 
\label{fig:combination}
}
\end{figure}

Fig.~\ref{fig:combination} summarizes our results showing both the $\gamma-$ray  and the neutrino fluxes for the same parameters of Fig. \ref{fig:UL_Cluster}.
Besides the  IceCube upper limits for clusters and diffuse neutrino background, it also depicts the diffuse $\gamma-$ray background (DGRB) data from Fermi-LAT~\citep{ackermann2015spectrum} and the upper limits from HAWC \citep{HAWC2022limits}. 
The $\gamma-$ray flux is given by the light-green, dark-green and olive-green bands, which are the counterparts of the light-blue, blue and dark-blue neutrino bands, respectively. 
We see that the olive-green, which is the most constrained band by the upper limits of the IceCube, falls below the Fermi data for $\gamma-$rays. 



\begin{figure}[ht!]
\centering
\includegraphics[width=0.9\textwidth]{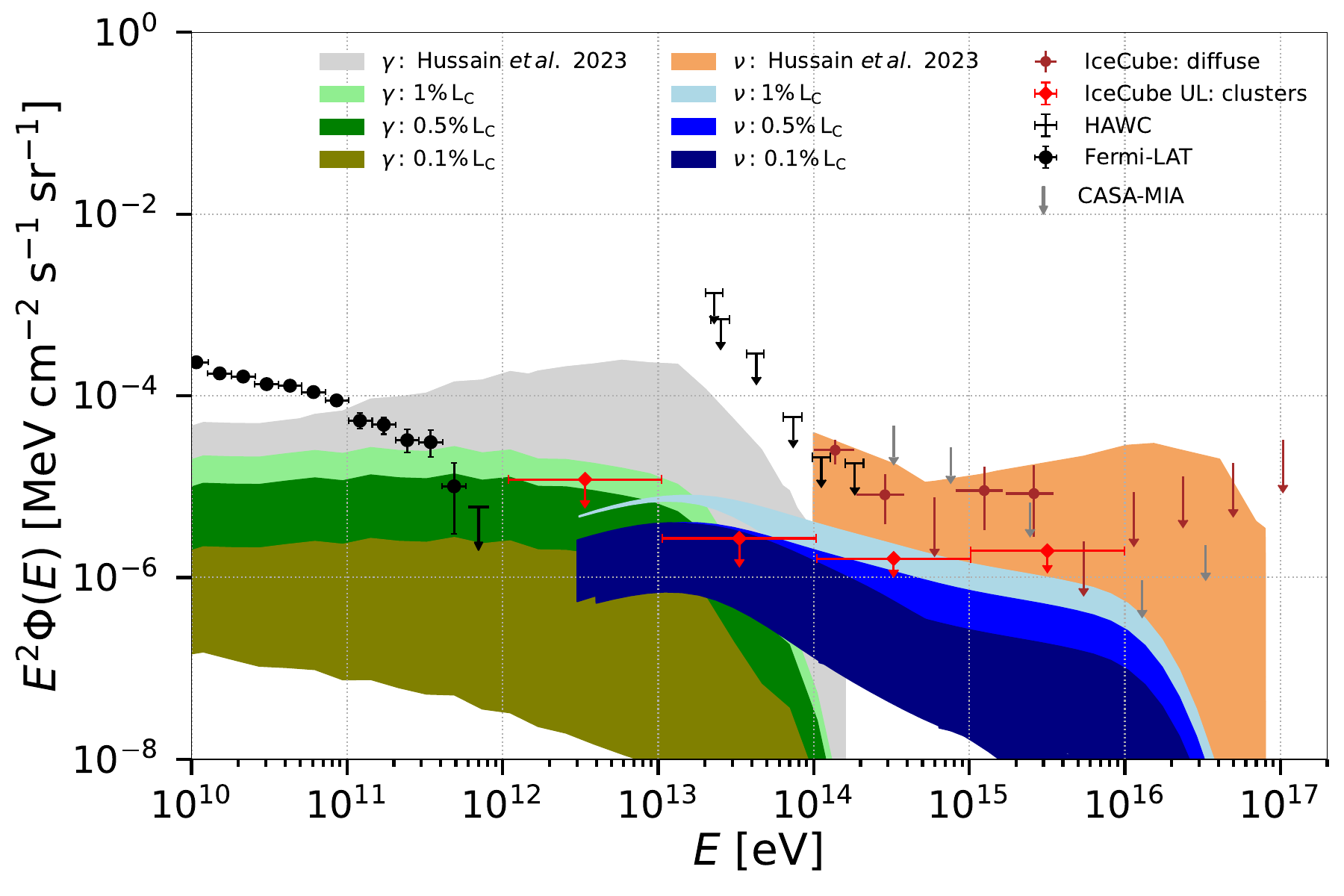}
\caption{ 
Neutrino (brown band) and $\gamma-$ray (gray band) fluxes form the entire population of galaxy clusters as obtained in \citep{hussain2023diffuse,hussain2021high}.
These are compared with the results of Figure \ref{fig:combination}, i.e., the flux of neutrinos (blue bands) obtained  with the new parametric space constrained by
the upper limits estimated by the IceCube \citep{abbasi2022searching}. The corresponding $\gamma-$ray flux (green bands) for the same parameters is also shown.
The diffuse neutrino background reported earlier by the IceCube \citep{aartsen2015searches, aartsen2015evidence} is also depicted.
The DGRB observed by Fermi-LAT~\citep{ackermann2015spectrum} and the upper limits for the DGRB from HAWC \citep{HAWC2022limits} and the CASA-MIA ~\citep{chantell1997limits} are also depicted. 
\label{fig:neuallUL}
}
\end{figure}

In Figure \ref{fig:neuallUL}  we compare the results obtained in Fig. \ref{fig:combination} with the earlier results of \citep{hussain2021high} and \citep{hussain2023diffuse}. We note that
the parametric space considered in these works cannot be excluded entirely by the upper limits of the IceCube \citep{abbasi2022searching}.

\begin{figure}[ht!]
\centering
\includegraphics[width=0.7\textwidth]{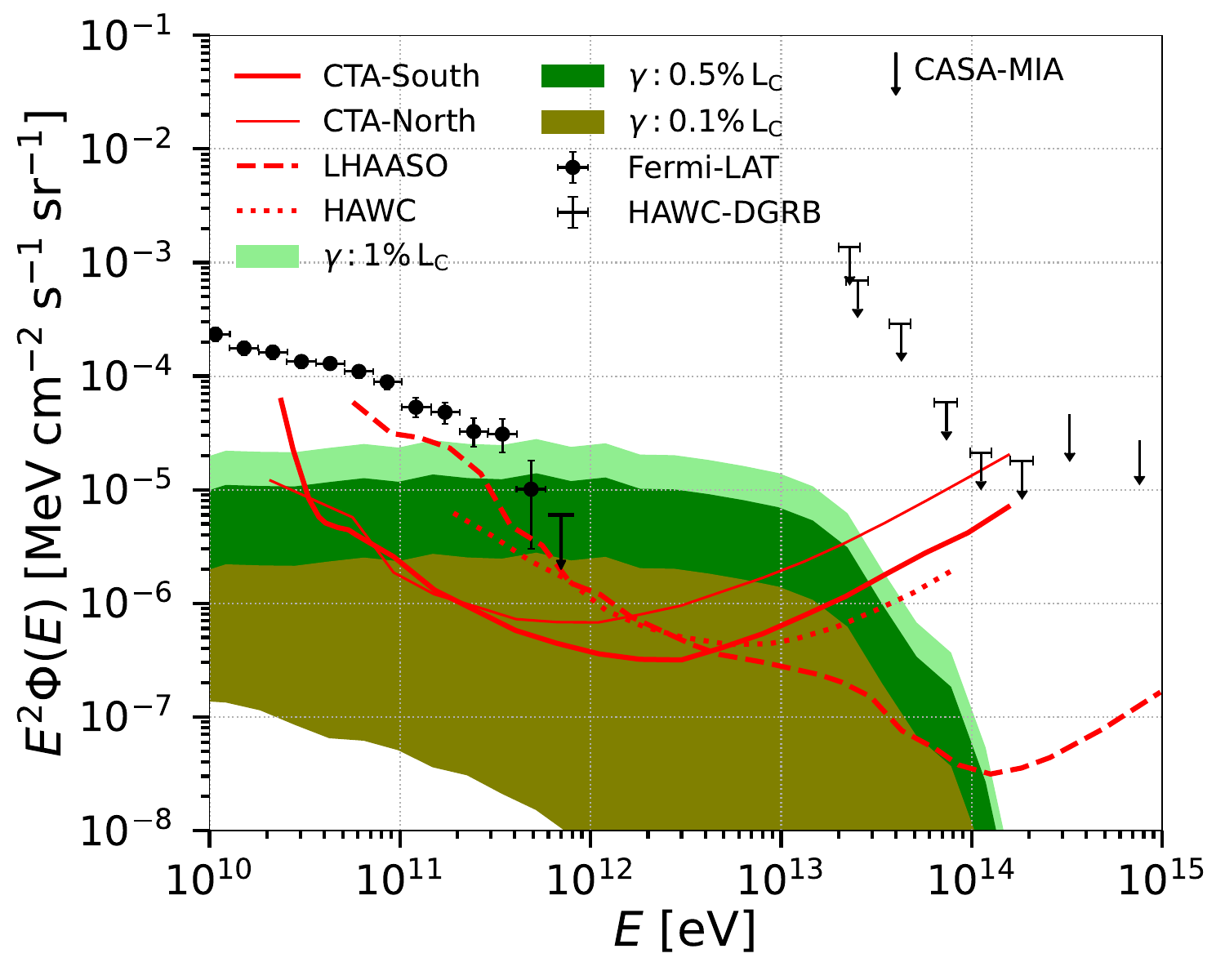}
\caption{
Flux of $\gamma-$rays  from the entire population of galaxy clusters as  in Figures \ref{fig:combination} and \ref{fig:neuallUL}. It is compared with sensitivity curves from different experiments, namely,
from Hawc for point-like sources \citep{abeysekara2013sensitivity}, LHAASO \citep{di2016lhaaso}, and CTA \citep{cta2018science}. Also depicted are the upper limits for DGRB from  HAWC \citep{HAWC2022limits} and  CASA-MIA \citep{CASAMIA1997limits}.
}
\label{fig:Sensitivity}
\end{figure}

In Fig. \ref{fig:Sensitivity}, we show the constrained $\gamma-$ray flux obtained in figure \ref{fig:combination}  for the entire population of clusters compared with sensitivity 
curves of different experiments. It includes the sensitivity curves for point-like sources from the High Altitude Water Cherenkov Observatory (HAWC)  \citep{abeysekara2013sensitivity}, Large High Altitude Air Shower Observatory (LHASSO) \citep{di2016lhaaso}, and the forthcoming  Cherenkov Telescope Array (CTA) \citep{cta2018science}, as well as the upper limits for the DGRB from HAWC \citep{HAWC2022limits} and CASA-MIA \citep{CASAMIA1997limits} experiments  \citep[see also][]{hussain2023diffuse}. Clearly, these observatories can potentially observe very high energy $\gamma-$rays from the clusters.

\section{Discussion and Conclusions}\label{sec:conclusion}

We have computed the flux of neutrinos and $\gamma-$rays from the entire population of galaxy clusters using the most detailed numerical simulations to date, considering 3D-MHD cosmological simulations of galaxy clusters up to redshift $z\lesssim 5.0$, as in \cite{hussain2021high, hussain2023diffuse}.  
According to these authors, clusters can contribute to a sizeable percentage of up to $100\%$ to diffuse neutrino background, depending on the parameters adopted for the CR spectrum. However,  
the IceCube collaboration \citep{abbasi2022searching} has reported that this contribution cannot exceed $(9-13)\%$. 
Evaluating upper limits for the neutrino flux of the clusters, this collaboration concluded that these new constraints would exclude the \cite{hussain2021high}  models for hard CR spectral indices $\alpha < 2.0$. Our present results indicate that, in fact, the new Icecube limits point to harder spectral indices $\alpha \geq 2.0$. Nevertheless, these new results are entirely compatible with the parametric space explored in \cite{hussain2021high}, which also included values of $\alpha \geq 2.0$ (see Figure \ref{fig:neuallUL}).

%
In particular, for $\alpha = 2.5$ and $E_\mathrm{max}=10^{17}$~eV the neutrino flux obtained in \citep{hussain2021high} is below the upper-limits of the IceCube and decreases approximately by an order of magnitude if we assume  CR luminosity $0.1\% \, L_C$ instead of $1\% \, L_C$.

We have also computed the $\gamma-$ray flux constrained by these IceCube upper limits and found that it decreases by roughly an order of magnitude in comparison with \cite{hussain2023diffuse}, for  CR luminosity $\sim 0.1\% \, L_C$. 
Despite that, the contribution of clusters to the DGRB is still substantial above $500$~GeV (Figure \ref{fig:neuallUL}). 
Moreover, the flux falls within the sensitivity ranges of the HAWC, LHAASO, and the upcoming CTA observatories, suggesting the possibility for direct detections of $\gamma-$rays from clusters of galaxies (Figure \ref{fig:Sensitivity}). 
As in the case of the neutrinos,  the $\gamma-$ray flux  goes for values lower than the Fermi-LAT observations \citep{ackermann2015spectrum} for $\alpha\sim 2.5$,  and reduces even more, if we consider a CR luminosity $0.1\% \, L_C$, rather than $1\% \,L_C$. While discrete source categories such as AGNs \citep{di2013diffuse, ajello2015origin} and star-forming galaxies \citep{roth2021diffuse} can make an important contribution to the DGRB for energy levels below the TeV range, our results highlight that the combined gamma-ray flux originating from clusters can surpass the combined impact of individual classes of unresolved sources for energies exceeding $500$~GeV. This aligns with the findings of  \cite{hussain2023diffuse}.

We should emphasize that our aim here was not to fit either IceCube upper limits \citep{abbasi2022searching} or Fermi-LAT data \citep{ackermann2015spectrum}. Instead, we have only compared our evaluation of the integrated flux of neutrinos and $\gamma-$ rays for the entire population of clusters with those observations.
Our estimations are dependent on the parametric space and so does the IceCube results.
%
To obtain their upper-limits,  \cite{abbasi2022searching} considered sources with masses in the range $10^{14} \leq M/M_{\odot} \leq 10^{15}$ up to redshift $z\leq 2.0$. Our analysis, on the other hand,  has considered the entire mass range of clusters $10^{12} \leq M/M_{\odot} \leq 2\times 10^{15}$, and redshifts $z\leq 5.0$.
Though major contribution to neutrino and $\gamma-$ray background comes from the nearby sources ($z\leq1$), and more massive clusters are more frequent,  the contribution from clusters in the mass range  $10^{13} \leq M/M_{\odot} \leq 10^{14}$  is not negligible \citep{hussain2021high, hussain2023diffuse}.  Therefore, 
including this mass interval might change the upper limits estimated by IceCube \citep{abbasi2022searching}.


Finally, we would like to stress that some specific parameters may have the potential to influence our simulations. For instance, a mixed composition of cosmic rays (CRs) and the distribution of CR sources within the clusters could lead to variations in our results. If we were to consider a CR composition involving heavy elements like iron (Fe), it could potentially alter our conclusions. Similarly, slight adjustments in the assumed distributions of CR sources within the structures (see Section \ref{sec:method})  might also have an impact on our outcomes. 
Exploring further these effects is a direction we plan to pursue in the future.
Furthermore, it is worth noting that we have not accounted for the influence of the uncertain diffuse 
magnetic field outside the clusters, which could introduce further minor variations in the gamma-ray observations.




\begin{acknowledgments}
We are indebt with Rafael Alves Batista for his very insighful comments and suggestions to this work.
 The work of SH and GP is partially supported by the research grant number $2017$W$4$HA$7$S ''NAT-NET: Neutrino and Astroparticle Theory Network'' under the program PRIN $2017$ funded by the Italian Ministero dell'Istruzione, dell'Universita' e della Ricerca (MIUR). 
 The work of EdGDP  is partially supported by the  Brazilian agencies FAPESP (grants $2013/10559-5$ and 
$ 2021/02120-0)$ and CNPq (grant  $308643/2017-8$). 
\end{acknowledgments}

\bibliographystyle{aasjournal}



\end{document}